\documentclass[showpacs,showkeys,pra]{revtex4}
\usepackage{epsf} \usepackage{color}
\usepackage{dcolumn}
\usepackage{bm}
\everymath{\displaystyle}
\begin{document} 

\title{Foldy-Wouthyusen Transformation and Semiclassical Limit for Relativistic
Particles in Strong External Fields}

\author{\firstname{Alexander J.}~\surname{Silenko}} \email{silenko@inp.minsk.by}
\affiliation{Institute of Nuclear Problems, Belarusian State University, Minsk 220030, Belarus}


\begin {abstract}
A general method of the Foldy-Wouthyusen (FW) transformation for
relativistic particles of arbitrary spin in strong external fields has been
developed. The use of the found transformation operator is not restricted
by any definite commutation
relations between even and odd operators. The final FW Hamiltonian
can be expanded into a power series in the Planck constant which
characterizes the order of magnitude of quantum corrections. Exact expressions for
low-order terms in the Planck constant can be derived. Finding
these expressions allows to perform a simple
transition to the semiclassical approximation which defines
a classical limit of the relativistic quantum mechanics.
As an example, interactions of spin-1/2 and scalar particles with
a strong electromagnetic field have been considered. Quantum and
semiclassical equations of motion of particles and their spins
have been deduced. Full agreement between quantum and classical
theories has been established.
\end{abstract}

\maketitle

\section{Introduction}

The Foldy-Wouthuysen (FW) representation \cite{FW} occupies a
special place in the quantum theory. Properties of this representation are
unique. The Hamiltonian and all
operators are block-diagonal (diagonal in two spinors). Relations
between the operators in the FW representation are similar to
those between the respective classical quantities. For
relativistic particles in external fields, operators
have the same form as in the nonrelativistic quantum theory. For
example, the position operator is $\bm r$ and the momentum one is
$\bm p=-i\hbar\nabla$.
These
properties considerably simplify the transition to the
semiclassical description. As a result, the FW representation
provides the best possibility of obtaining a meaningful classical
limit of the relativistic quantum mechanics. The basic advantages
of the FW representation are described in Refs.
\cite{FW,CMcK,JMP}.

Interactions of relativistic particles with strong external fields
can be considered on three levels: (i) classical physics, (ii)
relativistic quantum mechanics, and (iii) quantum field theory.
The investigation of such interactions on every level is
necessary. The use of the FW representation allows to describe
strong-field effects on level (ii) and to find a unambiguous
connection between classical physics and relativistic quantum
mechanics. To solve the problem, one should carry out an
appropriate FW transformation (transformation to the FW
representation). The deduced Hamiltonian should be exact up to
first-order terms in the Planck constant $\hbar$. This precision
is necessary for establishment of exact connection between the
classical physics and the relativistic quantum mechanics.

However, known methods of exact FW transformation either can be
used only for some definite classes of initial Hamiltonians in the
Dirac representation \cite{Nikitin,JMP} or need too cumbersome
derivations \cite{E}.


In the present work, new general method of the 
FW transformation for relativistic particles in strong external fields is 
proposed. This method gives exact expressions for low-order terms
in $\hbar$. The proposed method is based on the developments
performed in Ref. \cite{JMP} and can be utilized for particles of
arbitrary spin. Any definite commutation relations between even
and odd operators in the initial Hamiltonian are not needed. An
expansion of the FW Hamiltonian into a power series in the Planck
constant is used. Since just this constant defines the order of
magnitude of quantum corrections, the transition to the
semiclassical approximation becomes trivial. As an example,
interaction of scalar and spin-1/2 particles with a strong
electromagnetic field is considered.

We use the designations $[\dots,\dots]$ and $\{\dots,\dots\}$ for
commutators and anticommutators, respectively.



\section{Foldy-Wouthyusen transformation for 
particles in external fields}

In this section, we review previously developed methods of the FW transformation for 
particles in external fields.

The relativistic quantum mechanics is based on the Klein-Gordon equation
for scalar particles, the Dirac equation for spin-1/2 particles, and
corresponding relativistic wave equations for particles with higher spins
(see, e.g., Ref. \cite{FN}).
The quantum field theory is not based on these fundamental equations of
the relativistic quantum mechanics (see Ref. \cite{W}).
Relativistic wave equations for particles with any spin can be
presented in the Hamilton form. In this case, the Hamilton
operator acts on the bispinor wave function
$\Psi=\left(\begin{array}{c} \phi \\ \chi \end{array}\right)$:
\begin{equation} i\hbar\frac{\partial\Psi}{\partial t}={\cal
H}\Psi. \label{eqi} \end{equation} A particular case of Eq.
(\ref{eqi}) is the Dirac equation.

We can introduce the unit matrix $I$ and the 
Pauli matrices those components act on the spinors:
$$I=\left(\begin{array}{cc}1&0\\0&1\end{array}\right),
~~~\rho_1=\left(\begin{array}{cc}0&1\\1&0\end{array}\right),
~~~ \rho_2=\left(\begin{array}{cc}0&-i\\i&0\end{array}\right),
~~~ \rho_3\equiv\beta=\left(\begin{array}{cc}1&0\\0&-1\end{array}\right).$$

The Hamiltonian can be split into operators commuting and noncommuting with the operator $\beta$:
\begin{equation} {\cal H}=\beta {\cal M}+{\cal E}+{\cal O},~~~\beta{\cal M}={\cal M}\beta,
~~~\beta{\cal E}={\cal E}\beta, ~~~\beta{\cal O}=-{\cal O}\beta,
\label{eq3} \end{equation} where the operators ${\cal M}$ and
${\cal E}$ are even and the operator ${\cal O}$ is odd. We suppose
that the operator ${\cal E}$ is multiplied by the unit matrix $I$
which is everywhere omitted.

Explicit form of the Hamilton operators for particles with
arbitrary half-integer spin has been obtained in Ref. \cite{NGNN}.
Similar equations have been derived for spin-0 \cite{FV} and
spin-1 \cite{CS,YB} particles. To study semiclassical limits of
these equations, one should perform appropriate FW
transformations.

The wave function of a spin-1/2 particle can be transformed to a
new representation with the unitary operator $U$:
$$\Psi'=U\Psi.$$ The Hamilton operator in the new
representation takes the form \cite{FW,JMP,Gol} \begin{equation}
{\cal H}'=U{\cal H}U^{-1}-i\hbar U\frac{ \partial U^{-1}}{\partial
t}, \label{eq2}
\end{equation} or $$ {\cal H}'=U\left({\cal
H}-i\hbar\frac{\partial}{\partial t}\right)U^{-1}+
i\hbar\frac{\partial}{\partial t}. $$

The FW transformation has been justified in the best way. In the
classical work by Foldy and Wouthuysen \cite{FW}, the exact
transformation for free relativistic particles and the approximate
transformation for nonrelativistic particles in electromagnetic
fields have been carried out. There exist several other
nonrelativistic transformation methods which give the same results
(see Ref. \cite{JMP} and references therein). A few methods can be
applied for relativistic particles in external fields. However,
the transformation methods explained in Refs. \cite{B,GS} require
cumbersome calculations. 

The block-diagonalization of two-body Hamiltonians for a system of
two spin-1/2 particles and a system of spin-0 and spin-1/2
particles can be performed by the methods found by Chraplyvy
\cite{Chraplyvy} and Tanaka \emph{et al} \cite{Tanaka},
respectively.

Some methods allow to reach the FW representation without the use
of unitary transformations. The so-called elimination method
\cite{Pa} makes it possible to exclude the lower spinor from
relativistic wave equations. Variants of this method useful for
relativistic particles have been elaborated in Refs.
\cite{Neznamov,STMP}. Another method which essentially differs
from the FW and elimination methods has been presented in Ref.
\cite{Gos}. This method defines a diagonalization procedure based
on a formal expansion in powers of the Planck constant $\hbar$ and
can be used for a large class of Hamiltonians directly inducing
Berry phase corrections \cite{Gos}. An important feature of this
method is a possibility to take into account strong-field effects.

Any method different from the FW one should also be justified. The
validity of the elimination method is proved only by the
coincidence of results obtained by this method and the FW one
\cite{VJ}. Impressive agreement between results presented in Ref.
\cite{Gos} and corresponding results obtained by the FW method is
reached for first-order terms in $\hbar$. These terms define
momentum and spin dynamics that can be well described in the
framework of classical physics. To prove the validity of the
method, one should show such an agreement for terms derived from
second-order commutators (e.g., for the Darwin term \cite{FW}).
Therefore, we desist from a definitive estimate of the method
developed in Ref. \cite{Gos}.

We suppose the consistence with the genuine FW transformation to
be necessary for any diagonalization method. For example, the
Eriksen-Korlsrud method \cite{EK} does not transform the wave
function to the FW representation even for free particles
\cite{PRD}. The use of this method in Refs.
\cite{Obukhov,Heidenreich}
instead of the FW one could cause a
misunderstanding of the nature of spin-gravity coupling (see the
discussion in Refs. \cite{PRD,Mashhoon2}).

In the general case, the exact FW transformation has been found by
Eriksen \cite{E}. The validity of the Eriksen transformation has
also been argued by de Vries and Jonker \cite{VJ}. The Eriksen
transformation operator has the form \cite{E} \begin{equation}
U=\frac12(1+\beta\lambda)\left[1+\frac14(\beta\lambda+\lambda\beta-2)\right]^{-1/2},
~~~ \lambda=\frac{{\cal H}}{({\cal H}^2)^{1/2}}, \label{E}
\end{equation} where ${\cal H}$ is the Hamiltonian in the Dirac
representation. This operator brings the Dirac wave function and
the Dirac Hamiltonian to the FW representation in one step.
However, it is difficult to use the Eriksen method for obtaining
an explicit form of the relativistic FW Hamiltonian because the
general final formula is very cumbersome and contains roots of
Dirac matrix operators. Therefore, the Eriksen method was not used
for relativistic particles in external fields.

To perform the FW transformation in the strong external fields, we
develop the much simpler method elaborated in Ref. \cite{JMP} for
relativistic spin-1/2 particles. In this work, the initial Dirac
Hamiltonian is given by
\begin{equation} {\cal H}=\beta m+{\cal E}+{\cal O}, \label{eq3N} \end{equation}
where $m$ is the particle mass. In Eqs. (\ref{eq3N})--(\ref{eq31}), the system of units $\hbar=c=1$ is used.

When $[{\cal E},{\cal O}]=0$, the FW transformation is exact
\cite{JMP}. This transformation is fulfilled with the operator
\begin{equation} U=\frac{\epsilon+m+\beta{\cal
O}}{\sqrt{2\epsilon(\epsilon+m)}}, ~~~\epsilon=\sqrt{m^2+{\cal
O}^2} \label{eq18} \end{equation} and the transformed Hamiltonian
takes the form \begin{equation} {\cal H}_{FW}=\beta \epsilon+{\cal
E}. \label{eq17} \end{equation}

The same transformation is valid for Hamiltonian (\ref{eq3}) when
not only does the operator ${\cal E}$ commutates with ${\cal O}$
but also the operator ${\cal M}$:
\begin{equation} [{\cal M},{\cal O}]=0. \label{comm}
\end{equation} In this case, Eq. (\ref{eq17}) remains valid but
the operator $\epsilon$ takes the form
$$\epsilon=\sqrt{{\cal M}^2+{\cal O}^2}.$$

In the general case, the FW Hamiltonian has been obtained as a
power series in external field potentials and their derivatives
\cite{JMP}. As a result of the first stage of transformation
performed with operator (\ref{eq18}), the following Hamiltonian
can be found: \begin{equation} {\cal H}'=\beta\epsilon+{\cal
E}'+{\cal O}',~~~\beta{\cal E}'={\cal E}'\beta, ~~~\beta{\cal
O}'=-{\cal O}'\beta. \label{eq7}\end{equation} The odd operator
${\cal O}'$ is now comparatively small: \begin{equation}
\begin{array}{c} \epsilon=\sqrt{m^2+{\cal O}^2}, \\ {\cal
E}'=i\frac{\partial}{\partial t}+\frac{\epsilon+m}
{\sqrt{2\epsilon(\epsilon+m)}}\left({\cal
E}-i\frac{\partial}{\partial t}
\right)\frac{\epsilon+m}{\sqrt{2\epsilon(\epsilon+m)}}\\-\frac{\beta{\cal
O}} {\sqrt{2\epsilon(\epsilon+m)}}\left({\cal
E}-i\frac{\partial}{\partial t} \right)\frac{\beta{\cal
O}}{\sqrt{2\epsilon(\epsilon+m)}}, \\ {\cal O}'=\frac{\beta{\cal
O}}{\sqrt{2\epsilon(\epsilon+m)}} \left({\cal
E}-i\frac{\partial}{\partial t}
\right)\frac{\epsilon+m}{\sqrt{2\epsilon(\epsilon+m)}}\\-
\frac{\epsilon+m}{\sqrt{2\epsilon(\epsilon+m)}}\left({\cal
E}-i\frac{\partial} {\partial t}\right)\frac{\beta{\cal
O}}{\sqrt{2\epsilon(\epsilon+m)}}. \end{array} \label{eq28}
\end{equation}

The second stage of transformation leads to the approximate
equation for the FW Hamiltonian: \begin{equation} {\cal
H}_{FW}=\beta\epsilon+{\cal E}'+\frac14\beta\left\{{\cal
O}'^2,\frac{1}{\epsilon}\right\}. \label{eq31} \end{equation}

To reach a better precision, additional transformations can be used \cite{JMP}.

This method has been applied for deriving the Hamiltonian and the
quantum mechanical equations of momentum and spin motion for Dirac
particles interacting with electroweak \cite{JMP} and
gravitational \cite{PRD,PRD2} fields. The semiclassical limit of
these equations has been obtained \cite{JMP,PRD,PRD2}. To
determine the exact classical limit of the relativistic quantum
mechanics of arbitrary-spin particles in \emph{strong} external
fields, we need to generalize the method.

General properties of the Hamiltonian depend on the particle spin.
The Hamiltonian is hermitian (${\cal H}={\cal H}^\dagger$) for
spin-1/2 particles and pseudo-hermitian for spin-0 and spin-1 ones
(more precisely, $\beta$-pseudo-hermitian, see Ref.
\cite{Mostafazadeh} and references therein). In the latter case,
it possesses the property ($\beta^{-1}=\beta$) $${\cal
H}^\dagger=\beta{\cal H}\beta$$ that is equivalent to $${\cal
H}^\ddagger\equiv\beta{\cal H}^\dagger\beta={\cal H}.$$

The normalization of wave functions is given by
$$\int{\Psi^\dagger\Psi dV}=\int{(\phi\phi^\ast+\chi\chi^\ast)
dV}=1$$ for spin-1/2 particles and $$\int{\Psi^\ddagger\Psi
dV}\equiv\int{\Psi^\dagger\beta\Psi
dV}=\int{(\phi\phi^\ast-\chi\chi^\ast) dV}=1$$ for spin-0 and
spin-1 particles.

We suppose ${\cal M}={\cal M}^\dagger,~{\cal E}={\cal
E}^\dagger,~{\cal O}={\cal O}^\dagger$ when ${\cal H}={\cal
H}^\dagger$ and ${\cal M}={\cal M}^\ddagger,~{\cal E}={\cal
E}^\ddagger,~{\cal O}={\cal O}^\ddagger$ when ${\cal H}={\cal
H}^\ddagger$. These conditions can be satisfied in any case.

Since the FW Hamiltonian is block-diagonal and a lower spinor
describes negative-energy states, this spinor should be equal to
zero. The FW transformation should be performed with the unitary
operator $U^\dagger =U^{-1}$ for spin-1/2 particles and with the
pseudo-unitary operator $U^\ddagger\equiv\beta U^\dagger\beta
=U^{-1}$ for spin-0 and spin-1 particles.

\section {FOLDY-WOUTHYUSEN TRANSFORMATION IN STRONG EXTERNAL FIELDS}

We propose the method of the 
FW transformation for relativistic particles in strong external
fields which can be used for particles of arbitrary spin. The FW
Hamiltonian can be expanded into a power series in the Planck
constant which defines the order of magnitude of quantum
corrections. The obtained expressions for low-order terms in
$\hbar$ are exact. The proposed FW transformation makes the
transition to the semiclassical approximation to be trivial. The
power expansion can be available only if \begin{equation}
pl\gg\hbar, \label{rel1} \end{equation} where $p$ is the momentum
of the particle and $l$ is the characteristic size of the
nonuniformity region of the external field. This relation is
equivalent to \begin{equation} \lambda\ll l, \label{rel2}
\end{equation} where $\lambda$ is the de Broglie wavelength. Eqs.
(\ref{rel1}),(\ref{rel2}) result from the fact that the Planck
constant appears in the final Hamiltonian due to commutators
between the operators ${\cal M},{\cal E}$, and ${\cal O}$.

The expansion of the FW Hamiltonian into the power series in the
Planck constant is formally similar to the previously obtained
expansion \cite{JMP} into a power series in the external field
potentials and their derivatives. However, the equations derived
in Ref. \cite{JMP} do not define the semiclassical limit of the
Dirac equation for particles in strong external fields, while
these equations exhaustively describe the weak-field expansion.
The proposed method can also be used in the weak-field expansion
even when relations (\ref{rel1}),(\ref{rel2}) are not valid.

When the power series in the Planck constant is deduced, zero
power terms define the quantum analogue of the classical
Hamiltonian. On this level, classical and quantum expressions
should be very similar because the classical theory gives the
right limit of the quantum theory. Terms proportional to powers of
$\hbar$ may describe quantum corrections. As a rule, interactions
described by these terms also exist in the classical theory.
However, classical expressions may differ from the corresponding
quantum ones because the quantum corrections to the classical
theory may appear.


We generalize the method developed in Ref. \cite{JMP} in order to
take into account a possible non-commutativity of the operators
${\cal M}$ and ${\cal O}$. The natural generalization of
transformation operator (\ref{eq18}) used in Ref. \cite{JMP} is
\begin{equation} U=\frac{\beta\epsilon+\beta {\cal M}-{\cal
O}}{\sqrt{(\beta\epsilon+\beta {\cal M}-{\cal O})^2}}\,\beta,~~~
U^{-1}=\beta\,\frac{\beta\epsilon+\beta{\cal M}-{\cal
O}}{\sqrt{(\beta\epsilon+\beta{\cal M}-{\cal O})^2}},
\label{eq18N} \end{equation} where $U^\dagger=U^{-1}$ when ${\cal
H}={\cal H}^\dagger$, $U^{-1}=U^\ddagger$ when ${\cal H}={\cal
H}^\ddagger$, and $\epsilon=\sqrt{{\cal M}^2+{\cal O}^2}$. This
form of the transformation operator allows to perform the FW
transformation in the general case. The special case ${\cal
M}=mc^2$ has been considered in Ref. \cite{JMP} and commutation
relation (\ref{comm}) has been used in Refs. \cite{PRD,JETP}.

We consider the general case when external fields are
nonstationary. The exact formula for the transformed Hamiltonian
has the form
\begin{equation} \begin{array}{c} {\cal H}'=\beta\epsilon+{\cal
E}+ \frac{1}{2T}\Biggl(\left[T,\left[T,(\beta\epsilon+{\cal
F})\right]\right] +\beta\left[{\cal O},[{\cal O},{\cal
M}]\right]\\- \left[{\cal O},\left[{\cal O},{\cal F}\right]\right]
- \left[(\epsilon+{\cal M}),\left[(\epsilon+{\cal M}),{\cal
F}\right]\right] - \left[(\epsilon+{\cal M}),\left[{\cal M},{\cal
O}\right]\right]\\-\beta \left\{{\cal O},\left[(\epsilon+{\cal
M}),{\cal F}\right]\right\}+\beta \left\{(\epsilon+{\cal
M}),\left[{\cal O},{\cal F}\right]\right\} \Biggr)\frac{1}{T},
\end{array} \label{eq28N} \end{equation} where
${\cal F}={\cal E}-i\hbar\frac{\partial}{\partial t}$ and
$T=\sqrt{(\beta\epsilon+\beta{\cal M}-{\cal O})^2}$.

Hamiltonian (\ref{eq28N}) still contains odd terms proportional to
the first and higher powers of the Planck constant. This
Hamiltonian can be presented in the form \begin{equation} {\cal
H}'=\beta\epsilon+{\cal E}'+{\cal O}',~~~\beta{\cal E}'={\cal
E}'\beta, ~~~\beta{\cal O}'=-{\cal O}'\beta,
\label{eq27}\end{equation} where $\epsilon=\sqrt{{\cal M}^2+{\cal
O}^2}.$ The even and odd parts of Hamiltonian (\ref{eq27}) are
defined by the well-known relations: $${\cal
E}'=\frac12\left({\cal H}'+\beta{\cal
H}'\beta\right)-\beta\epsilon,~~~ {\cal O}'=\frac12\left({\cal
H}'-\beta{\cal H}'\beta\right).$$

Additional transformations performed according to Ref. \cite{JMP}
bring ${\cal H}'$ to the block-diagonal form. The approximate
formula for the final FW Hamiltonian is \begin{equation} {\cal
H}_{FW}=\beta\epsilon+{\cal E}'+\frac14\beta\left\{{\cal
O}'^2,\frac{1}{\epsilon}\right\}. \label{eqf} \end{equation} This
formula is similar to the corresponding one obtained in Ref.
\cite{JMP}. The additional transformations allow to obtain more
precise expression for the FW Hamiltonian.

Eqs. (\ref{eq28N})--(\ref{eqf}) solve the problem of the FW
transformation for relativistic particles of arbitrary spin in
strong external fields.

Eq. (\ref{eq28N}) can be significantly simplified in some special
cases. When $[{\cal M},{\cal O}]=0$ and the external fields are
stationary, it is reduced to
\begin{equation} \begin{array}{c} {\cal H}'=\beta\epsilon+{\cal
E}+ \frac{1}{2T}\Biggl(\left[T,\left[T,{\cal E}\right]\right] \\-
\left[{\cal O},\left[{\cal O},{\cal E}\right]\right] -
\left[(\epsilon+{\cal M}),\left[(\epsilon+{\cal M}),{\cal
E}\right]\right] \\-\beta \left\{{\cal O},\left[(\epsilon+{\cal
M}),{\cal E}\right]\right\}+\beta \left\{(\epsilon+{\cal
M}),\left[{\cal O},{\cal E}\right]\right\} \Biggr)\frac{1}{T}.
\end{array} \label{eqrd} \end{equation} In this case,
$[\epsilon,{\cal M}]=[\epsilon,{\cal O}]=0$ and the operator
$T=\sqrt{2\epsilon(\epsilon+{\cal M})}$ is even.

\section {SPIN-1/2 AND SCALAR PARTICLES IN STRONG ELECTROMAGNETIC FIELD}

As an example, the FW transformation for spin-1/2 and
scalar particles interacting with a strong electromagnetic field
can be considered.
The initial Dirac-Pauli Hamiltonian for a particle possessing an
anomalous magnetic moment (AMM) has the form \cite{P}
\begin{equation}\begin{array}{c} {\cal
H}_{DP}=c\bm{\alpha}\cdot\bm{\pi}+\beta mc^2+e\Phi+\mu'(-\bm
{\Pi}\cdot \bm{H}+i\bm{\gamma}\cdot\bm{ E}),\\
\bm{\pi}=\bm{p}-\frac{e}{c}\bm{ A}, ~~~
\mu'=\frac{g-2}{2}\cdot\frac{e\hbar}{2mc}, \end{array}
\label{eqDP} \end{equation} where $\mu'$ is the AMM, $\Phi,\bm{
A}$ and $\bm{ E},\bm{H}$ are the potentials and strengths of the
electromagnetic field.

Here and below the following designations for the matrices are used:
$$\begin{array}{c}\bm{\gamma}=\left(\begin{array}{cc} 0  &  \bm{\sigma} \\ -\bm{\sigma} & 0
\end{array}\right), ~~~ {\beta}\equiv\gamma^0=\left(\begin{array}{cc} 1  &  0
\\ 0 & -1 \end{array}\right), ~~~\bm{\alpha}=\beta\bm\gamma=
\left(\begin{array}{cc} 0  &  \bm{\sigma} \\ \bm{\sigma} & 0
\end{array}\right), \\    \bm{\Sigma}
=\left(\begin{array}{cc} \bm{\sigma}  &  0 \\ 0 &
\bm{\sigma}\end{array}\right),   ~~~\bm{\Pi}=\beta\bm\Sigma
=\left(\begin{array}{cc} \bm{\sigma}  &  0 \\ 0 &
-\bm{\sigma}\end{array}\right),  \end{array}  $$
where $0,1,-1$ mean the corresponding 2$\times$2 matrices and $\bm{\sigma}$ is
the Pauli matrix.

Terms describing the electric dipole moment (EDM) $d$ have been
added in Ref. \cite{RPJ}. The resulting Hamiltonian is given by
\begin{equation}\begin{array}{c} {\cal
H}=c\bm{\alpha}\cdot\bm{\pi}+\beta mc^2+e\Phi+\mu'(-\bm {\Pi}\cdot
\bm{H}+i\bm{\gamma}\cdot\bm{ E}) 
-d(\bm {\Pi}\cdot
\bm{E}+i\bm{\gamma}\cdot\bm{ H}), ~~~
d=\frac{\eta}{2}\cdot\frac{e\hbar}{2mc}, \end{array} \label{eqEDM}
\end{equation} where $\eta$ factor for the EDM is an analogue of
$g$ factor for the magnetic moment.
It is important that $\mu'$ and $d$ are proportional to $\hbar$.

In the considered case $$\begin{array}{c} {\cal M}=mc^2,~~~ {\cal
E}=e\Phi-\mu'\bm {\Pi}\cdot \bm{H}-d\bm {\Pi}\cdot \bm{E}, ~~~
{\cal O}=c\bm{\alpha}\cdot\bm{\pi}+i\mu'\bm{\gamma}\cdot\bm{
E}-id\bm{\gamma}\cdot\bm{ H}. \end{array}$$

Since only terms of zero and first powers in the Planck constant
define the semiclassical equations of motion of particles and
their spins, we retain only such terms in the FW Hamiltonian. The
terms of order of $\hbar$ are proportional either to field
gradients or to products of field strengths ($H^2,~E^2$ and $EH$).
We do not calculate the terms proportional to products of field
strengths because they are usually small in comparison with the
terms proportional to field gradients.

The calculated Hamiltonian is given by
\begin{equation}\begin{array}{c} {\cal
H}_{FW}=\beta\epsilon'+e\Phi- \mu'\bm\Pi\cdot\bm
H-\frac{\mu_0}{2}\left\{\frac{mc^2}{\epsilon'}, \bm\Pi\cdot\bm
H\right\}\\+\frac{\mu'c}{4}\left\{\frac{1}{\epsilon'},
\left[\bm\Sigma\cdot(\bm \pi\times \bm E)-\bm\Sigma\cdot(\bm
E\times \bm\pi) \right]\right\}\\+
\frac{\mu_0mc^3}{\sqrt{2\epsilon'(\epsilon'+mc^2)}}\left[\bm\Sigma\cdot(\bm
\pi\times \bm E)-\bm\Sigma\cdot(\bm E\times \bm\pi)
\right]\frac{1}{\sqrt{2\epsilon'(\epsilon'+mc^2)}}\\+
\frac{\mu'c^2}{2\sqrt{2\epsilon'(\epsilon'+mc^2)}}\left\{(\bm{\Pi}\cdot\bm\pi),
(\bm{H}\cdot\bm\pi+\bm{\pi}\cdot\bm
H)\right\}\frac{1}{\sqrt{2\epsilon'(\epsilon'+mc^2)}}\\
-d\bm\Pi\cdot\bm E 
+\frac{dc^2}{2\sqrt{2\epsilon'(\epsilon'+mc^2)}}\left\{(\bm{\Pi}\cdot\bm\pi),
(\bm{E}\cdot\bm\pi+\bm{\pi}\cdot\bm
E)\right\}\frac{1}{\sqrt{2\epsilon'(\epsilon'+mc^2)}}\\
-\frac{dc}{4}\left\{\frac{1}{\epsilon'}, \left[\bm\Sigma\cdot(\bm
\pi\times \bm H)-\bm\Sigma\cdot(\bm H\times \bm\pi)
\right]\right\}, \end{array} \label{eq33} \end{equation} where
\begin{equation}
\epsilon'=\sqrt{m^2c^4+c^2\bm{\pi}^2} \label{eq34} \end{equation}
and $\mu_0=\frac{e\hbar}{2mc}$ is the Dirac magnetic moment.

The quantum evolution of the kinetic momentum operator, $\bm\pi$,
is defined by the operator equation of particle motion:
\begin{equation}\frac{d\bm\pi}{dt}=\frac{i}{\hbar}[{\cal
H}_{FW},\bm\pi] -\frac{e}{c}\cdot\frac{\partial\bm A}{\partial t}.
\label{eqme} \end{equation}

The equation of spin motion describes the evolution of the
polarization operator $\bm\Pi$:
\begin{equation}\frac{d\bm\Pi}{dt}=\frac{i}{\hbar}[{\cal
H}_{FW},\bm\Pi]. \label{eqpoe} \end{equation}

Because the operator $\bm\pi$ does not contain the Dirac spin
matrices, the commutator of this operator with the Hamiltonian is
proportional to $\hbar$.
The equation of spin-1/2 particle motion in the strong
electromagnetic field to within first-order terms in the Planck
constant has the form \begin{equation} \begin{array}{c} \frac{d\bm
\pi}{dt}=e\bm E+\beta\frac{ec}{4}\left\{\frac{1}{\epsilon'},
\left([\bm\pi\times\bm H]-[\bm H\times\bm\pi]\right)\right\}
+\mu' \nabla(\bm\Pi\cdot\bm H)+
\frac{\mu_0}{2}\left\{\frac{mc^2}{\epsilon'},
\nabla(\bm\Pi\cdot\bm
H)\right\}\\-\frac{\mu'c}{4}\left\{\frac{1}{\epsilon'},
\left[\nabla(\bm\Sigma\cdot[\bm\pi\times\bm E])-
\nabla(\bm\Sigma\cdot[\bm E\times\bm\pi]) \right]\right\}\\-
\frac{\mu_0mc^3}{\sqrt{2\epsilon'(\epsilon'+mc^2)}}\left[\nabla(\bm\Sigma\cdot[\bm\pi\times\bm
E])- \nabla(\bm\Sigma\cdot[\bm
E\times\bm\pi])\right]\frac{1}{\sqrt{2\epsilon'(\epsilon'+mc^2)}}\\
-\frac{\mu'c^2}{2\sqrt{2\epsilon'(\epsilon'+mc^2)}}\left\{(\bm{\Pi}\cdot\bm\pi),\left[\nabla
(\bm{H}\cdot\bm\pi)+\nabla(\bm{\pi}\cdot\bm H)\right]\right\}
\frac{1}{\sqrt{2\epsilon'(\epsilon'+mc^2)}}. \end{array}
\label{eq35} \end{equation}

This equation can be divided into two parts. The first part does
not contain the Planck constant and describes the quantum
equivalent of the Lorentz force. The second part is of order of
$\hbar$. This part defines the relativistic expression for the
Stern-Gerlach force. Since terms proportional to $d$ are small,
they are omitted.

The equation of spin motion is given by
\begin{equation} \begin{array}{c} \frac{d\bm{\Pi}}{dt}=\frac{2\mu'}{\hbar}\bm\Sigma\times\bm H+\frac{\mu_0}{\hbar} \left\{\frac{mc^2}{\epsilon'},
\bm\Sigma\times\bm H\right\} 
-\frac{\mu'c}{2\hbar}\left\{\frac{1}{\epsilon'},
\left[\bm\Pi\times(\bm \pi\times \bm E)-\bm\Pi\times(\bm E\times \bm\pi) \right]\right\}\\
-
\frac{\mu_0mc^3}{\hbar\sqrt{\epsilon'(\epsilon'+mc^2)}}\left[\bm\Pi\times(\bm
\pi\times \bm E)
-\bm\Pi\times(\bm E\times \bm\pi) \right]\frac{1}{\sqrt{\epsilon'(\epsilon'+mc^2)}}\\
- \frac{\mu'c^2}{\hbar\sqrt{2\epsilon'(\epsilon'+mc^2)}}
\left\{(\bm\Sigma\times\bm \pi),
(\bm{H}\cdot\bm\pi+\bm{\pi}\cdot\bm H)\right\}
\frac{1}{\sqrt{2\epsilon'(\epsilon'+mc^2)}} \\
+\frac{2d}{\hbar}\bm\Sigma\times\bm E 
-\frac{dc^2}{\hbar\sqrt{2\epsilon'(\epsilon'+mc^2)}}
\left\{(\bm\Sigma\times\bm \pi),
(\bm{E}\cdot\bm\pi+\bm{\pi}\cdot\bm E)\right\}
\frac{1}{\sqrt{2\epsilon'(\epsilon'+mc^2)}}
\\+\frac{dc}{2\hbar}\left\{\frac{1}{\epsilon'}, \left[\bm\Pi\times(\bm
\pi\times \bm H)-\bm\Pi\times(\bm H\times \bm\pi) \right]\right\}.
\end{array} \label{eq36} \end{equation}

Eqs. (\ref{eq33}),(\ref{eq35}),(\ref{eq36}) agree with the corresponding equations derived in Refs. \cite{JMP,RPJ}.
However, unlike the latter equations, Eqs. (\ref{eq33}),(\ref{eq35}),(\ref{eq36}) describe strong-field effects.


We can also consider the interaction of spinless particles with
the strong electromagnetic field. The initial Klein-Gordon
equation describing this interaction has been transformed to the
Hamilton form in Ref. \cite{FV}.

In this case, the Hamiltonian acts on the two-component wave
function which is the analogue of the spinor. The explicit form of
this Hamiltonian is \cite{FV} \begin{equation} {\cal
H}=\rho_3mc^2+(\rho_3+i\rho_2)\frac{\bm\pi^2}{2m}+e\Phi.
\label{FV} \end{equation} Therefore, \begin{equation} {\cal
M}=mc^2+\frac{\bm\pi^2}{2m}, ~~~{\cal E}=e\Phi, ~~~{\cal
O}=i\rho_2\frac{\bm\pi^2}{2m}, ~~~[{\cal M},{\cal O}]=0.
\label{MEO} \end{equation} For spinless particles,
\begin{equation} \epsilon=\sqrt{m^2c^4+c^2\bm\pi^2}, ~~~
T=\sqrt{\frac{\epsilon}{mc^2}}\left(\epsilon+mc^2\right).
\label{eqep} \end{equation}

The Hamiltonian transformed to the FW representation is given by
\begin{equation} {\cal H}_{FW}=\beta\epsilon+{\cal
E}=\beta\sqrt{m^2c^4+c^2\bm\pi^2}+e\Phi. \label{eqfz}
\end{equation} There are not any terms of order of $\hbar$ in this
Hamiltonian, while it contains terms of second and higher orders
in the Planck constant. We do not calculate the latter terms
because their contribution into equations of particle motion is
usually negligible.

The operator equation of particle motion takes the form
\begin{equation} \begin{array}{c} \frac{d\bm \pi}{dt}=e\bm
E+\beta\frac{ec}{4}\left\{\frac{1}{\epsilon},
\left([\bm\pi\times\bm H]-[\bm H\times\bm\pi]\right)\right\}.
\end{array} \label{eqpz} \end{equation}

The right hand side of this equation coincides with the
spin-independent part of the corresponding equation for spin-1/2
particles.

Eq. (\ref{eqfz}) for the FW Hamiltonian agrees with Eq.
(12) in Ref. \cite{PAN}. In this reference, the weak-field approximation
has been used and the operator equation of particle motion in the strong
electromagnetic field has not been obtained.

\section { SEMICLASSICAL LIMIT OF RELATIVISTIC QUANTUM MECHANICS FOR PARTICLES IN STRONG 
EXTERNAL FIELDS}

To obtain the semiclassical limit of the relativistic quantum
mechanics, one needs to average the operators in the quantum
mechanical equations. When the FW representation is used and
relations (\ref{rel1}),(\ref{rel2}) are valid, the semiclassical
transition consists in trivial replacing operators by
corresponding classical quantities. In this representation, the
problem of extracting even parts of the operators does not appear.
Therefore, the derivation of equations for particles of arbitrary
spin in strong external fields made in the precedent section
solves the problem of obtaining the semiclassical limit of the
relativistic quantum mechanics. If the momentum and position
operators are chosen to be the dynamical variables, relations
(\ref{rel1}),(\ref{rel2}) are equivalent to the condition
\begin{equation} |<p_i>|\cdot|<x_i>|\gg|<[p_i,x_i]>|=\hbar,~~~
i=1,2,3. \label{rel3} \end{equation} The angular brackets which
designate averaging in time will be hereinafter omitted.

Obtained semiclassical equations may differ from corresponding
classical ones.

As a result of replacing operators by corresponding classical
quantities, the semiclassical equations of motion of particles and
their spins take the form \begin{equation} \begin{array}{c}
\frac{d\bm \pi}{dt}=e\bm E+\frac{ec}{\epsilon'}
\left(\bm\pi\times\bm H\right)
+\mu'\nabla(\bm P\cdot\bm H)+ \frac{\mu_0}{mc^2\epsilon'} \nabla(\bm P\cdot\bm H)\\
-\frac{\mu'c}{\epsilon'} \nabla(\bm P\cdot[\bm\pi\times\bm E])
-\frac{\mu_0mc^3}{\epsilon'(\epsilon'+mc^2)}\nabla(\bm
P\cdot[\bm\pi\times\bm E])\\
-\frac{\mu'c^2}{\epsilon'(\epsilon'+mc^2)}(\bm{P}\cdot\bm\pi)\nabla
(\bm{H}\cdot\bm\pi), ~~~~~~~ \bm P=\frac{\bm S}{S}, \end{array}
\label{eqw} \end{equation}
\begin{equation} \begin{array}{c} \frac{d\bm P}{dt}=2\mu'\bm P\times\bm H+ \frac{2\mu_0mc^2}{\epsilon'}( \bm P\times\bm H)
-\frac{2\mu'c}{\epsilon'} \left(\bm P\times[\bm \pi\times \bm
E]\right)\\ -\frac{2\mu_0mc^3}{\epsilon'(\epsilon'+mc^2)}\left(\bm
P\times[\bm \pi\times \bm E] \right)
-\frac{2\mu'c^2}{\epsilon'(\epsilon'+mc^2)} (\bm P\times\bm \pi)(\bm{\pi}\cdot\bm H)\\ %
+2d\bm P\times\bm E-\frac{2dc^2}{\epsilon'(\epsilon'+mc^2)} (\bm
P\times\bm \pi)(\bm{\pi}\cdot\bm E) 
+\frac{2dc}{\epsilon'} \left(\bm P\times[\bm \pi\times \bm
H]\right). \end{array} \label{eqt} \end{equation} In Eqs.
(\ref{eqw}),(\ref{eqt}), $\epsilon'$ is defined by Eq.
(\ref{eq34}), $\bm P$ is the polarization vector, and $\bm S$ is
the spin vector (i.e., the average spin).

For scalar particles
\begin{equation} \begin{array}{c}
\frac{d\bm \pi}{dt}=e\bm E+\frac{ec}{\sqrt{m^2c^4+c^2\bm\pi^2}}
\left(\bm\pi\times\bm H\right). \end{array} \label{eqwl}
\end{equation}

Two first terms in right hand sides of Eqs.
(\ref{eqw}),(\ref{eqwl}) are the same as in the classical
expression for the Lorentz force. This is a manifestation of the
correspondence principle. The part of Eq. (\ref{eqt}) dependent on
the magnetic moment coincides with the well-known
Thomas-Barg\-mann-Michel-Te\-leg\-di (T-BMT) equation. It is
natural because the T-BMT equation has been derived without the
assumption that the external fields are weak. The relativistic
formula for the Stern-Gerlach force can be obtained from the
Lagrangian consistent with the T-BMT equation (see Ref.
\cite{PK}). The semiclassical and classical formulae describing
this force also coincide. High-order corrections to the quantum
equations of motion of particles and their spins should bring a
difference between quantum and classical approaches.

\section {DISCUSSION AND SUMMARY}

The new method of the FW transformation for relativistic particles
of arbitrary spin in strong external fields described in the
present work is based on the previous developments \cite{JMP}.
However, the use of transformation operator (\ref{eq18N}) is not
restricted by any definite commutation relations [see Eq.
(\ref{comm})] between even and odd operators. The proposed method
utilizes the expansion of the FW Hamiltonian into a power series
in the Planck constant which defines the order of magnitude of
quantum corrections. In the FW Hamiltonian, exact expressions for
low-order terms in $\hbar$ can be obtained. If the de Broglie
wavelength is much less than the characteristic size of the
nonuniformity region of the external field [see Eqs.
(\ref{rel1}),(\ref{rel2})], the transition to the semiclassical
approximation becomes trivial. In this case, it consists in
replacing operators by corresponding classical quantities. The
simplest semiclassical transition is one of main preferences of
the FW representation.

If Eqs. (\ref{rel1}),(\ref{rel2}) are not valid, the proposed
method can be used in the weak-field expansion. This expansion
previously used in Ref. \cite{JMP} presents the FW Hamiltonian as
a power series in the external field potentials and their
derivatives. In this case, the operator equations characterizing
dynamics of the particle momentum and spin can also be 
derived. Solutions of these equations define the quantum evolution
of main operators. Semiclassical evolution of classical quantities
corresponding to these operators can be obtained by averaging the
operators in the solutions. An example of such an evolution is
time dependence of average energy and momentum in a two-level
system.

When the FW Hamiltonian can be expanded into a power series in the
Planck constant, we obtain the semiclassical limit of the
relativistic quantum mechanics. Since the correspondence principle
must be satisfied, classical and semiclassical Hamiltonians and
equations of motion must agree. As an example, we consider the
interaction of scalar and spin-1/2 particles with the strong
electromagnetic field. We have carried out the FW transformation
and have derived the quantum equations of particle motion. We have
also deduced the quantum equations of spin motion for spin-1/2
particles. Averaging operators in the quantum equations consists
in substitution of classical quantities for these operators and
allows to obtain the semiclassical equations which are in full
agreement with the corresponding classical equations. The proved
agreement confirms the validity of both the correspondence
principle and the aforesaid method. All calculations have been
carried out for relativistic particles in strong external fields.

\section*{Acknowledgements}

The author is grateful to O.V. Teryaev for interest in the work
and helpful discussions and to T. Tanaka for bringing to his
attention Ref. \cite{Tanaka}. This work was supported by the
Belarusian Republican Foundation for Fundamental Research.

 \end{document}